\documentstyle[draft,12pt]{article}

\begin{document}

\newcommand{\dslash}{\mbox{ $\not \!\! D$}}

\newcommand{\unmezzo} {\frac{1}{2}}
\newcommand{\unterzo} {\frac{1}{3}}
\newcommand{\unquarto}{\frac{1}{4}}

\newcommand{\parzialet}[1]   {  \frac{ \partial{#1} }{\partial{t}} }
\newcommand{\parzialexi}[1]  {  \frac{ \partial{#1} }{\partial{x_i}} }
\newcommand{\parzialexxi}[1] {  \frac{ {\partial}^2{#1} }{\partial {x_i^2}} }
\newcommand{\derivatax}[1] {  \frac{        d{#1} }{        dx} }
\newcommand{\derivatat}[1] {  \frac{        d{#1} }{        dt} }

\newcommand{\traccia}{ {\rm Tr} }

\newcommand{\be}{\begin{equation}}
\newcommand{\ee}{\end{equation}}
\newcommand{\ca}{ {\cal A} }
\newcommand{\ch}{ {\cal H} }
\newcommand{\cm}{ {\cal M} }

\title{The Phase Diagram of Fluid Random Surfaces with Extrinsic Curvature}

\author{
Mark BOWICK, Paul CODDINGTON, Leping HAN,\\
Geoffrey HARRIS and Enzo MARINARI$^{(*)}$\\[1.5em]
Dept. of Physics and NPAC,\\
Syracuse University,\\
Syracuse, NY 13244, USA\\
{\footnotesize
  bowick@suhep.phy.syr.edu
  paulc@sccs.syr.edu
  han@sccs.syr.edu}\\
{\footnotesize
  gharris@sccs.syr.edu
  marinari@roma2.infn.it}\\[1.0em]
{\small $(*)$:  and Dipartimento di Fisica and INFN,} \\
{\small Universit\`a di Roma {\it Tor Vergata}}\\
{\small Viale della Ricerca Scientifica, 00173 Roma, Italy}}
\maketitle

\vfill
\begin{abstract}
We present the results of a large-scale simulation of a
Dynamically Triangulated Random Surface with extrinsic curvature
embedded in three-dimensional flat space.
We measure a variety of local observables and use
a finite size scaling analysis to characterize as much as
possible the regime of crossover from crumpled to smooth surfaces.
\end{abstract}
\vfill

\begin{flushright}
  {\bf ROM2F-92-48}\\
  {\bf SU-HEP-4241-517}\\
  {\bf SCCS-357}\\
  {\bf hep-lat/9209020 }\\
\end{flushright}

\newpage

\section{Introduction \protect\label{S_INT} }

In this paper we use Monte Carlo simulations to investigate a
theory of bosonic strings embedded in three target space dimensions
with the addition of an extrinsic curvature term to the action.
We present a complete high-statistics analysis of the behaviour of
a set of relevant observables. Since computing correlation functions
on dynamically triangulated surfaces
is a difficult task,
we have focused on elucidating the phase diagram
by analyzing local observables in great detail.

String theory, in a number of guises, has been conjectured to describe the
underlying fundamental physics of a wide variety of physical phenomena
and models. These include the strong interaction at long distances,
the three-dimensional Ising model and unified models incorporating
gravity.  In its  simplest form, the
bosonic string, it is a theory of free fluctuating surfaces.  The
functional integral for the Euclideanized bosonic string is just the
partition function for an ensemble of  random fluctuating fluid surfaces.
Such surfaces are also ubiquitous in nature, being found for example
in macro-emulsions
and the lipid bilayers that form an important part of cell
membranes \cite{DAVREV}.
These systems are fluid because
their component `molecules' are loosely bound. Their constituents are
arranged so that the net surface tension (nearly) vanishes; thus these
membranes are subject to large thermal fluctuations.   In one important
respect, however, these chemical/biological membranes differ
fundamentally from the
surfaces we discuss and simulate; they are self-avoiding.
The worldsheets of the bosonic string, in contrast,
generically self-intersect.

The bosonic string for surfaces embedded in $26$ dimensional space has been
studied extensively.  Much progress in numerically simulating strings has
been made through the use of
Dynamically Triangulated Random Surfaces (DTRS)\cite{DYTRAS,KAZMIG}.
For theories of surfaces embedded in $ D \leq 1$,
analysis of the continuum Liouville theory \cite{KPZ,DADIKA} and
of matrix models has led to
consistent predictions for critical exponents and correlation functions.
In the `double scaling limit', in fact, exact
nonperturbative solutions \cite{EXACT} have been found from these matrix
models; the functional integral over surfaces has been essentially
summed over all genera.

These analytical
techniques have failed, for the most part, in probing the theory of
random surfaces in the more interesting embedding regime $D > 1$.
Indeed the
formulas for critical exponents computed in Liouville theory give
complex results when continued to the range $1<D<25$.\footnote{More
precisely, one encounters these instabilities in Liouville theory when the
quantity $c - 24\Delta  >1$, where $c$ denotes the central charge of the matter
theory which describes the embedding of the surfaces and $\Delta$ is the
conformal weight of the lowest weight state in this theory \cite{KUTSEI}.
Since here we are considering flat space, $c=D$ and $\Delta=0$.} The matrix
models describing $D>1$ strings have so far been too complicated to solve.

Monte Carlo simulations for $D>1$
\cite{BILDAV,AMNOEX,EALRYM,MINOEX} indicate that these theories do
not appear to describe the fluctuations of two dimensional {\em smooth}
surfaces in the continuum limit.  Extremely spiky, branched-polymer-like
configurations with high (perhaps infinite) Hausdorff dimension dominate the
functional integral\footnote{In the same way singular configurations
dominate the Gaussian theory, which is essentially a theory of free random
walks.}.

Evidence for this pathology was obtained, for instance, when it was shown
that the bare string tension, essentially the amount
of work per unit area needed to perturb a boundary loop on these
configurations, cannot vanish at the critical point \cite{AMNEED}. (This
implies that the renormalized string tension diverges in the continuum
limit.) It has been
speculated in ref. \cite{CATES} that the proliferation of vortex
configurations of the internal geometry (the `Liouville mode') induces
the degeneration of these surfaces in the embedding space.  A related
explanation that is often proffered is that  a negative mass-squared
particle\footnote{In some contexts this is referred to as the `tachyon'.},
which comes on shell in the string theory for $D >1$, creates instabilities
which are made manifest by these singular configurations.

The tachyon, and apparently these related instabilities, can be eliminated in
particular cases by introducing fermionic coordinates and supersymmetry on
the worldsheet, and implementing an appropriate projection of states.
Presumably, the fermions effectively smooth out the
surfaces.  This would be consistent with what has been observed for
one-dimensional geometries; the random walk of a spin one-half particle has
Hausdorff dimension one and thus appears to be smooth \cite{POLBOO}. Many
authors have proposed an alternative modification of the string action
\cite{CANHAM,HELFRI,POLEXT,KLEINA} via the addition of a term that directly
suppresses extrinsic curvature\footnote{In fact, integrating the fermions out
of the Green-Schwarz superstring yields an action similar to the one we
consider, but with the addition of a complex Wess-Zumino type term
\cite{WIEGMA}.}. We shall examine this class of theories in this paper.

To write down our action we introduce an explicit parametrization of a
generic surface $\cm$ in $R^{3}$ with coordinates $(\sigma _1, \sigma _2)$
and the embedding $X^{\mu}(\sigma _i)$. $\mu$ runs from $1$ to $3$ (since
we only study the case of a $3d$ embedding space).  The induced metric
(the pullback of the  Euclidean $R^{3}$ metric via the embedding) is given by

\be
  h_{ij} = \partial _{\sigma_i} X^{\mu}\partial _{\sigma_j} X_{\mu}\ .
\ee

We will use Greek letters for the embedding space indices; they can be raised
and lowered at will since our background space is flat. Associated with each
point in $\cm$ are tangent vectors  ($t^{\mu}_i {\in} T\cm$) and a normal
vector  $n^{\mu} {\in} T\cm^{\perp}$.  The extrinsic curvature matrix $K_{ij}$
(the second fundamental form) can be defined by

\begin{equation}
\protect\label{kijdef}
	 \partial_{i} n^{\mu} = - K_{ij}{t^{\mu}}^j \ .
\end{equation}
The eigenvalues of this matrix are the inverses of the radii of curvature
of $\cm$. One usually describes the geometry of
these surfaces in terms of the mean curvature \cite{CHOBRU,DOCARM}

\begin{equation}
            H = \frac{1}{2}h^{ij}K_{ij}\ ,
\end{equation}
and the Gaussian curvature

\begin{equation}
           K = \epsilon^{ik}\epsilon^{jl}K_{ij}K_{kl}\ .
\end{equation}

One can show that the Gaussian curvature can be computed solely from
the metric $h_{ij}$, while the mean curvature depends explicitly on
the embedding $X^{\mu}$.

        We shall be concerned primarily with the Polyakov form of the
string action \cite{POLSTR}, in which an additional  {\em intrinsic
metric} $g_{ij}$ is introduced to describe the surface geometry.  We
discretize our model by triangulating surfaces.  In this construction,
each triangle is equilateral with area $1$ in the intrinsic  metric; the
coordination number at each vertex determines the intrinsic curvature of the
surface.  The coordinates $i$ label the vertices of the triangulation. Then
the discrete analogue of the intrinsic metric is the adjacency matrix
$C_{ij}$ whose elements equal $1$ if $i$ and $j$ label neighbouring nodes of
the triangulation, and vanish otherwise.  Two-dimensional diffeomorphism
invariance reduces to the permutation symmetry of the adjacency matrix at
this discrete level. One of the keys, in fact, to the power of this
construction is the preservation of this symmetry. Each vertex of the
triangulation is embedded in $R^3$ via the mapping $X^{\mu}_i$.  Given the
embedding $X$, we can also associate a unit normal vector
$(n^{\mu})_{\dot{k}} $ with each triangle on the surface (dotted Roman
indices label the triangles).  Note that all of the surface curvature of our
triangulations is concentrated along the links and vertices. The
surface is still flat  in the direction tangent (but not transverse) to each
link, so that the mean curvature has support on the links, while the Gaussian
curvature is non-zero only at the vertices.
The intrinsic curvature $R_i$ at vertex $i$ is
given by the deficit angle determined solely by the triangulation

\begin{equation}
     	R_i = \pi\frac{(6 - q_i)}{q_i},
\end{equation}
where $q_i$ denotes the connectivity of the lattice at vertex $i$.
The Gaussian curvature $K$ on the other hand is expressed in terms of
the deficit angle in the embedding space.

We shall study the theory defined by the action

\begin{equation}
	 S = S_{Gauss} + \lambda S_E =
  \sum_{i,j,\mu}C_{ij}(X^{\mu}_i - X^{\mu}_j)^2 +
  \lambda\sum_{\dot{k},\dot{l},\mu}C^{\dot{k}\dot{l}}(1 -
  n^{\mu}_{\dot{k}}\cdot n^{\mu}_{\dot{l}})\ .
\end{equation}
Thus, for $\lambda > 0$, we have introduced a ferromagnetic interaction
in the surface normals.
The model defined by this action has been studied
in \cite{CATTER,BAJOWI,BCJW,AMBONE,AMBTWO,CAKORE} and references therein.

{}From (\ref{kijdef}) and the definition of the induced metric, it follows that
this is a discretization of the continuum action

\begin{equation}
  \protect\label{ourcontact}
  S = \int\sqrt{|{\det {g}}|}(g^{ij}\partial_i X^{\mu}
  \partial_j X^{\mu}  + \frac{\lambda}{2}g^{ij}h^{kl}K_{ik}K_{jl})\ .
\end{equation}

Note that the second term in this action is manifestly positive, Weyl and
reparametrization invariant, and that $\lambda$ is a dimensionless coupling.
So, naively, it is not clear whether it is relevant or not. If it were
relevant, one would then anticipate that (since it obeys all of the
appropriate symmetries) it should be effectively generated in any string
action, and that it should engender ordering of the normals. It should then
lead to another RG fixed point at a finite value of $\lambda$, which would
characterize a phase transition between the crumpled phase (observed when
$\lambda=0$) and a `smooth(er)' phase.\footnote{The extrinsic curvature
term is also higher-derivative, indicating that the field theory described
by this action is non-unitary. This fact alone does not imply that the
associated
string-scattering amplitudes do not satisfy unitarity. Polchinski and Yang
\cite{POLYAN} do, however, contend that in this case the string theory will not
be unitary.  Even if this were so, this model could still
be an appropriate description of the statistical mechanics of fluctuating
surfaces, although not one corresponding to a
physical fundamental string theory.}
We proceed first to review previous work which has addressed the question of
whether or not these theories exhibit a crumpling transition.

\subsection{Previous Analytical Work}

A renormalization group analysis \cite{HELFRI,KLEINB,FORSTE,POLEXT} indicates,
however, that there should be no phase transition at finite coupling when
such extrinsic curvature dependent operators are added to the action.  The
computations of refs. \cite{HELFRI,KLEINB,FORSTE} use the action

\begin{equation}
  \protect\label{acthkf}
     S = \int d^2\sigma\ (\mu_0\sqrt{\det{h}} + \frac{1}{\alpha}
    \sqrt{\det{h}}(h^{ij}K_{ij})^2)\ ,
\end{equation}
in the regime in which the string tension $\mu_0$ is small (unlike the usual
particle physics limit of string theory, which is characterized by large
$\mu_0$). After integrating out fluctuations of the embedding $X^{\mu}$
between momentum scales $\Lambda$ and $\tilde{\Lambda}$, it is found that the
renormalization of the extrinsic curvature coupling is given to one-loop
order by

\begin{equation}
  \beta ({\alpha}) \equiv \Lambda \frac{d\alpha}{d\Lambda}
  = - \frac{3}{4\pi}\alpha ^2 \ ,
\end{equation}
so that $\alpha$ is driven to infinity in the infra-red.
This theory thus exhibits asymptotic freedom. Surfaces are smooth
(the normals are correlated) below a persistence length \cite{PELLEI}

\begin{equation}
  \protect\label{persist}
  \xi_p \sim \exp(\frac{4\pi}{3\alpha_{bare}})\ ,
\end{equation}
and are disordered above this scale.  Some
intuition into this result can be gained by observing that this theory is
similar to the $O(3)$ sigma model, which is  asymptotically free
\cite{POLBOO}. The normals to $M$ are the analogues of $O(3)$ vectors, though
in this case they are constrained to be normal to a surface governed by the
action (\ref{ourcontact}).

Without the extrinsic curvature term, (\ref{acthkf}) is the Nambu-Goto
action, while (\ref{ourcontact}), which we use in our simulations,  is based
on the action quantized by Polyakov.  Classically (when the equations of
motion for the Polyakov action are solved and substituted back into the
action), the two actions are equivalent.  It has also been demonstrated
\cite{MORRIS} that the two quantizations are equivalent in the
critical dimension $D=26$.  In
lower dimensions (note that the Nambu-Goto action clearly does not make sense
for $D < 2$), it is not so clear that quantizations `based' on the two
actions are indeed the same.  The work of Polchinski and Strominger
\cite{POLSTRO} suggests that there are alternative quantizations.
Distler (\cite{DIST3D}) has also questioned the equivalence of these
quantizations in $D=3$.  Indeed, even if the two quantizations are
equivalent, it
does not automatically follow that the two theories are still the same once
an extrinsic curvature dependent term has been added.

In fact, Polyakov in \cite{POLEXT} uses a hybrid form of the
action (\ref{acthkf}) and still
obtains the same result for the beta function. He introduces an intrinsic
metric $g_{ij}$, chooses the conformal gauge $g_{ij}=\rho \delta_{ij}$ and
considers

\begin{equation}
  \protect\label{polyact}
  S = \frac{1}{2\alpha}\int d^2\sigma( \mu_o \rho + \rho^{-1}
  (\partial^2X^{\mu})(\partial^2X^{\mu}) + \lambda^{ij}(\partial_iX^{\mu}
  \partial_jX^{\mu} - \rho\delta_{ij}))\ .
\end{equation}

Classically, the Lagrange multiplier $\lambda^{ij}$ constrains the intrinsic
metric to equal the induced metric (this equality is not enforced by the
classical equations of motion for the original Polyakov action).  This
constraint should be relaxed quantum mechanically if, as Polyakov
\cite{POLBOO} argues,  the condensate of this Lagrange multiplier assumes a
value of the order of the momentum cutoff.  If this dynamical assumption is
correct, then one can essentially derive the equivalence of this Nambu-Goto
like  and the original Polyakov quantizations.  In the large $D$ (embedding
dimension) limit, saddle point calculations \cite{DAGUAE} show that $\lambda$
indeed does acquire a large expectation value, and that for small values of
the string tension $\mu_o$, the coupling $\alpha$ is asymptotically free, as
the RG calculations suggest.

There are, however, a couple of caveats and suggestions in the analytic
literature that do allow for the existence of a crumpling transition for
fluid surfaces.   Polyakov remarks that if, in the infrared region,
fluctuations of the internal geometry ($\rho$) are suppressed relative to
fluctuations of the extrinsic metric, then the beta function is proportional
to $\alpha$ and hence the continuum limit of the theory exhibits non-trivial
scaling behaviour; this presumably cannot be the case in the large $D$
limit.  Another RG calculation, performed by Yang \cite{YANG} using the
Polchinski-Strominger action \cite{POLSTRO} with an extrinsic curvature
dependent term, indicates that the two-loop correction (which is proportional
to $\alpha^3$) might be large enough to yield a zero of the beta
function, and thus a non-trivial IR fixed point.  The Polchinski-Strominger
action is  based on the assumption that the Liouville mode $\rho$ effectively
decouples (its mass is much greater than the momentum scale set by the
string tension); it is not clear why this assumption should hold for the
model that we simulate.  Finally, note that these computations are
perturbative (in $1/D$ or $\alpha$). It is possible that
non-perturbative effects could drive a crumpling transition.

\subsection{Previous Numerical Evidence}

     Monte Carlo simulations of the action (\ref{ourcontact}) on dynamically
triangulated random surfaces (DTRS) were first performed by Catterall
\cite{CATTER}, and  shortly thereafter by Baillie, Johnston, and Williams
\cite{BAJOWI,BCJW} and Catterall, Kogut and Renken \cite{CAKORE}.  They
simulated triangulations with the topology of the sphere, and measured the
specific heat

\be
  C(\lambda) \equiv \frac{\lambda^2}{N}(<S_E^2> - <S_E>^2)\ ,
\ee
on surfaces with up to $N = 144$ nodes (and $N = 288$ nodes in the last
reference).  They found a peak in the specific heat; the peak size appeared
to grow with $N$.  A similar model that can be vectorized
rather straightforwardly was also considered; the set of planar
$\phi^3$ graphs was
simulated \cite{CAKORE,CEKR}.  Each vertex of these $\phi^3$ graph was
embedded in $R^3$ and the action (\ref{ourcontact}) was used; graphs of up to
$1000$ nodes were simulated (these would be dual to $500$ node
triangulations).  It was found that the specific heat peak grew with $N$,
albeit slowly, as

\begin{equation}
\label{specfit}
            C_{max} = AN^{\omega} + B\ ,
\end{equation}
with $\omega = 0.185(50)$.  Further work by Ambj{\o}rn, Irb\"ack, Jurkiewicz,
Petersson and Varsted \cite{AMBONE,AMBTWO}, using dynamical triangulations
with the topology of the torus and lattices with up to $N=576$ nodes,
indicated that the rate of increase of the peak height severely diminishes
with increasing $N$.  The  data strongly suggests
that in fact the specific heat peak height does {\bf not} diverge as
$N\rightarrow\infty$.  These authors also measured the bare string tension
and mass gap, by embedding the torus in a background toroidal space spanned
by a loop, and measuring the dependence of the free energy on the loop size.
They found that these measurements (when taken for $\lambda$ values near the
peak position) are consistent with the appropriate scaling relations (with
vanishing bare string tension and mass gap) that should characterize a phase
transition to smooth surfaces.
This measurement, although it constitutes the best evidence there
is so far for a real phase transition at $\lambda=\lambda_c$,
is still quite an indirect way of measuring correlation functions.  As we
will discuss, these scaling relations could contradict other
observed phenomena such as the absence of diverging correlation
times and increasing finite size
effects at the putative critical point.

Thus it appears that numerical evidence could allow for the existence of a
crumpling transition (most probably of higher order), while analytical
calculations generally indicate that no such transition should occur.

In  \cite{KROGOM} the peak was measured in a DTRS simulation that
incorporated self-avoidance and the extrinsic curvature term $S_E$, with  a
solid-wall potential substituted for the Gaussian term in the action. The
results for the specific heat turned out to be very similar to those found in
the simulations we have just discussed, for example, in \cite{AMBTWO}.  The
specific heat peak is, in this context, considered to be a lattice
artifact, because the peak height levels off with large $N$ (of order $500$).
These simulations included a crude block-spin measurement that suggests that
the renormalization group flow of $\lambda$ is consistent with the analytical
result of asymptotic freedom.

Simulations using other discretizations for the extrinsic curvature dependent
term have yielded somewhat different results  \cite{CATTER,BAJOWI}.  The
specific heat peak, measured in simulations employing  what is referred to as
the `area discretization', is rather feeble, and levels off for small
values of $N$ (by $N=72$) (the authors interpret this as being indicative of
perhaps a `third' order transition). Actions based on these various
discretizations have been simulated for fixed, triangular meshes.  These
systems model tethered or crystalline membranes, in which the constituent
molecules are tightly bound together.  In the tethered case, the specific heat
peak obtained from simulations of the edge action (\ref{ourcontact}) grows
vigorously as a function of $N$ for very large ($128 \times 128$) lattices
\cite{WHEATE}.  This is strong evidence for the existence  of a second order
transition which, in this case, is in accord with the analytic results --
these
calculations are reviewed by Nelson \cite{NELSON} and David
\cite{DAVREV,DAVREZ} and involve mean field and large $D$ computations which
suggest that the $\beta$ function is linear at leading order, with a zero for
finite $\alpha$, i.e. a UV fixed point.   When the alternate area
discretization is used in the tethered case, the specific heat peak again
stops growing.  Recent work has demonstrated that this other discretization is
pathological in the tethered case; the  class of `corrugated' surfaces,
which are singular in one direction and smooth in the other, then dominates
the path integral \cite{WHEATE}.

Thus, given the muddle of somewhat contradictory evidence,  it is unclear
whether or not a crumpling transition exists for fluid surfaces.  We have
pursued this question by taking high statistics measurements of the specific
heat peak, and by measuring other observables describing the geometry of these
surfaces, since observables with different quantum numbers
can give quite different information.
For example, in the Ising model the magnetization behaves quite unlike
the internal energy (which is invariant under the standard
$Z_2$ transformation).

To analyze and interpret this data, we have applied insights
gained from work on better understood systems, primarily spin models and
lattice gauge theories. Issues of the equivalence of the
Nambu-Goto and Polyakov quantizations have also motivated us to compare the
intrinsic and induced geometry of the surfaces that we simulate.

\subsection{The Plan of the Paper}

We hope that in this section we have introduced the problem in sufficient
detail. In Section \ref{S_OBS} we define the quantities we have decided to
measure, and explain why they are physically interesting. Next, in
Section \ref{S_NUM}, we present
the details of our numerical simulations. In Section \ref{S_PHA} we describe
and discuss our results concerning the phase diagram of the theory and we
devote Section \ref{S_TIM} to the discussion of correlation times. In
Section \ref{S_RES} we propose various interpretations of these
results and in Section \ref{S_WHA}
we comment on future possible developments.

\section{Observables \protect\label{S_OBS} }

To minimize finite size effects, we have considered triangulations with the
topology of the torus.  The action (\ref{ourcontact}) was used, with
the BRST invariant measure utilized also by Baillie, Johnston, and Williams
(\cite{BAJOWI}), so that

\begin{equation}
  Z = \sum_{G \in T(1)}
  \int\prod_{\mu,i}dX^{\mu}_{i}\prod_{i}q_i^{\frac{d}{2}}
  \exp( -S_{Gauss} - \lambda S_E)\ ,
\end{equation}
where $d=3$, $q_i$ is the connectivity of the $i$th vertex, and $T(1)$ refers
to the set of triangulations of genus $1$. The authors of
\cite{AMBONE,AMBTWO} do not include this connectivity dependent term in
their measure.  The long-distance physics of the simulations is
presumably insensitive to the presence of this term.
Because we have chosen a
different measure, though, our quantitative results cannot be
precisely compared with theirs.

     We measured a variety of quantities that characterize the extrinsic and
intrinsic geometry of these surfaces.  These observables include:

\begin{enumerate}

\item

  The edge curvature $S_E$ and the associated specific heat $C(\lambda)$,
  which is a sensitive indicator of the presence of a phase
  transition.

\item

  The squared radius of gyration $R_G$;

  \begin{equation}
    R_G \equiv \frac{1}{N}\sum_{i,\mu}(X_i^{\mu} - X_{{\rm{com}}}^{\mu})^2\ ,
    \protect\label{E_RG}
  \end{equation}
  where the com subscript refers to the center of mass of the surface.
  By measuring the $N$ dependence of the gyration radius, we can extract
  a value for the extrinsic Hausdorff dimension, which is given by

  \begin{equation}
    R_G \sim N^{\nu}\sim N^{\frac{2}{d_{\rm{extr}}}}\ \ .
    \protect\label{E_RGD}
  \end{equation}

\item

  The magnitude of the extrinsic Gaussian curvature. We measure
  a discretization of $\int \mid K \mid \sqrt{\mid h \mid}$,
  with

  \begin{equation}
    \mid {\cal{K}} \mid = \frac{1}{N}\sum_i\mid 2\pi -
    \sum_{\dot{j}} \phi_i^{\dot{j}}\mid\ .
    \protect\label{E_CK}
  \end{equation}
  Here $\phi_i^{\dot{j}}$ denotes the angle subtended by the $\dot{j}$th
  triangle at the $i$th vertex.  This quantity, therefore, measures the
  magnitude of the deficit angle in the embedding space averaged over
  all vertices. We also record the mean square fluctuation of
  $\mid{\cal{K}}\mid$, denoted by $F[\mid{\cal{K}}\mid]$.

\item

  The corresponding intrinsic quantity, $\mid {\cal{R}} \mid$, given by

  \begin{equation}
    \mid {\cal{R}} \mid = \frac{\pi}{3N}\sum_i\mid 6 - q_i \mid\ ,
    \protect\label{E_CR}
  \end{equation}
  and its fluctuations.  When the intrinsic and extrinsic metrics are equal,
  the intrinsic and extrinsic deficit angles are identical, and
  $K = R/2$.

\item

  To study the correlation between intrinsic and extrinsic geometry,
  we also measure the quantity which we refer to as ${\cal{K}}*{\cal{R}}$:

  \begin{equation}
    {\cal{K}}*{\cal{R}} \equiv \frac{\int KR}{\sqrt{\int K^2 \int R^2}} =
    \frac{\sum_i(2\pi - \sum_{\dot{j}}\phi_i^{\dot{j}})(6 - q_i)}{\sqrt{
    \sum_i(2\pi - \sum_{\dot{j}}\phi_i^{\dot{j}})^2\sum_i(6 - q_i)^2}}\ .
    \protect\label{E_KR}
  \end{equation}

  This quantity is $1$ when the metrics are equal, $0$ if they are un-
  correlated, and negative when these curvatures are anti-correlated.

\item

  We measure, finally, the average maximum coordination number of the
  surface vertices,  ${\rm{max}}_i \ q_i$.

\end{enumerate}

\section{The Numerical Simulation \protect\label{S_NUM} }

In our simulations we have used the standard Metropolis algorithm to update
the embedding fields $X^\mu_i$.  To sweep through the space of
triangulations we performed flips (see reference \cite{EALRYM}) on randomly
chosen links.  Flips were automatically rejected if they yielded a degenerate
triangulation; i.e. one in which a particular vertex has fewer than three
neighbors, or in which a vertex is labeled as its own neighbor, or where more
than one link connects two vertices. (It has been proven in ref.
\cite{EALRYM,MINOEX} that the entire space of graphs of a given topology can
be spanned by only performing these flips.)  After a set of  $3N$ flips was
performed, $3N$ randomly selected embedding coordinates were updated via
random shifts from a flat distribution,

\be
  X^{\mu} \rightarrow X^{\mu} + \delta X^{\mu}\ .
\ee

The mean magnitude of these shifts

\be
  <\delta X^{\mu} \delta X^{\mu}>
\ee
was chosen so that the acceptance rate for updates of the $X^{\mu}$ was
roughly $50$ percent.  Most of the  Monte Carlo simulations were performed on
HP-9000 (720 and 750 series) workstations; we also collected some data by
simulating lattices on each of the 32 nodes of a CM-5.  Our code was in
Fortran, with a Fibonacci random number generator.

\begin{table}
\begin{tabular}{|l||l|l|l|l|l|l|l|l|} \hline
N=144 &.8 &1.25 & 1.35 & 1.40 & 1.45 & 1.50 & 2.0 & $\lambda$\\ \cline{2-8}
 & 3 & 3 & 3 & 3 & 3 & 3 & 3 & $\times 10^6$ sweeps \\ \hline
N=288 & .8 & & 1.375 & 1.40 & 1.425 & 1.475 & 2.0 & $\lambda$\\ \cline{2-8}
 & 14.4 && 21.0 & 15.0 & 16.2 & 13.5 & 14.4 & $\times 10^6$ sweeps \\ \hline
N= 576 & .8 & 1.325 & 1.375 & 1.40 & 1.425 & 1.475 & 2.0 & $\lambda$
\\ \cline{2-8}
& 12.0 & 27.0 & 27.0 & 27.0 & 27.0 & 27.0 & 9.6 & $\times 10^6$ sweeps \\
\hline
\end{tabular}
\protect\caption[CT_ONE]{A record of the number of sweeps performed
at each different $\lambda$ value for $3$ different lattice sizes.
\protect\label{T_ONE}}
\end{table}

In Table (\ref{T_ONE}) we summarize our runs. Note that we have
performed quite
long runs on the larger lattice sizes. We will discuss in Section
(\ref{S_TIM}) why we believe runs of this length are just sufficient to yield
accurate values of the observables for the largest lattice size ($N=576$).

In all of our figures the different dots will be printed with their associated
statistical error (sometimes too small to be visible). The statistical error
is computed by means of a standard binning procedure. We will explicitly
discuss the cases in which our estimator for the statistical error is not
asymptotic.

The lines in these figures are from a histogram reconstruction (see for example
\cite{ENZOHI,SWEFEA}). We patch different histograms
\cite{SWEFEB,ALBESA,SWESTA} by weighting them with the associated statistical
indetermination (which we estimate by a jack-knife binned procedure); this
procedure seems to be very effective and reliable. All of the reconstruction
curve sets ($3$: dotted, dashed and continuous for $3$ surface sizes on each
figure) consist of $3$ curves (which sometimes appear as a single
one). The middle curve is the  histogram reconstruction, and the upper and the
lower ones bound the data within the errors obtained by the
procedure we have just described.

For $N=144$ we have patched the four histograms originating from
$\lambda=1.35$, $1.40$, $1.45$, $1.50$. For $N=288$ we have used
$\lambda=1.375$, $1.40$, $1.425$ and $1.475$. For $N=576$, we chose
$\lambda=1.375$,
$1.40$ and $1.425$.

We have only drawn the reconstructed, patched curves (with their reliable
errors) in the regions where we trust them. For example, close to the
pseudo-critical region we can trust a peak pattern only when we can
reconstruct the peak by starting from both sides of the transition (without
multi-histogram patching). So we have always used single histogram
reconstructions to check these criteria, before constructing the final,
multi-histogram data.

\section{The Phase Diagram \protect\label{S_PHA} }

We have measured, as stated previously, a large number of local
observables. We will see that a mixed picture emerges from these
measurements. For example the observables related to the dynamical
triangulations exhibit a characteristic pattern,
to be discussed in detail below.

We start by showing, in Fig.1, the edge curvature $S_E$ as a function of
$\lambda$.  The crossover region is around $\lambda \simeq 1.4$. For small
values of $\lambda$, the surface is crumpled (see the latter part of
this section). In this region,
finite size effects are already negligible for our lattice sizes, and our $3$
data points are on top of each other. We can see weak finite size effects by
comparing the continuous lines in the transient region. The $N=144$ dotted
line is far from the ones of the two larger lattices, which lie, on the
contrary, on top of each other. Finite size effects are larger in the large
$\lambda$ phase. One would expect, close to a phase transition with a
diverging correlation length, an increase of the finite size effects which we
do not observe here. The lattice should feel the presence of the zero mass
excitation, and the finite size corrections could be larger than everywhere
else (in the case of periodic boundary conditions they would obey a
power-law, rather than decaying
exponentially with size). This is surely not firm evidence
against the presence of a phase transition, but it does show that the
putative critical behaviour is atypical.

The errors in the `flat phase' ($\lambda=2.0$) are not under control. Our
estimators do not plateau under repeated iterations of the binning
procedure.  In this regime, correlation times are
large, as we will discuss in next section. This {\em caveat} holds for this
figure and for all the quantities we have measured.

In Fig. 2 we show the related specific heat $C(\lambda)$, in the same
$\lambda$ region. In Fig. 3 we enlarge the pseudo-critical $\lambda$ region,
in order to show the reconstructed peak of the specific heat. As already
noted our reconstruction procedure is quite reliable here.

In Table \ref{T_TWO} we give the maximum of the edge curvature specific heat
and its location for the $3$ different lattice sizes.

\begin{table}
\begin{tabular}{|l|l|l|} \hline
N & $C(\lambda)^{\rm{max}}$ & $\lambda_c$ \\ \hline
144 & 5.37(14) & 1.395(30) \\ \hline
288 & 5.55(7) & 1.410(25) \\ \hline
576 & 5.81(17) & 1.425(30) \\ \hline
\end{tabular}
\protect\caption[CT_TWO]{The maximum of the specific heat and its position,
with their errors, for $3$ different lattice sizes. \protect\label{T_TWO}}
\end{table}

We can extract from these data a specific heat exponent
$\omega = .06\pm .05$, with $\omega$ defined as in equation  (\ref{specfit}),
and the constant $B$ set to zero.  If we estimate an effective exponent from
the two smaller lattice sizes we get $.05\pm.06$, and from the two larger ones
we get $.07\pm.06$; this demonstrates that we do not see, within our
statistical precision, any
sign of a non-pure-power, non-asymptotic behavior. Note that, if
the constraint that $B$ vanishes is relaxed, our data is not
accurate enough to yield a meaningful fit to equation (\ref{specfit}).
A very small
(asymptotically finite) correlation length is sufficient to produce
such a small effect on our quite small lattice sizes. These results appear to
be consistent with those of the Copenhagen group \cite{AMBTWO}, and they are
not so far from the ones of the Urbana group  \cite{CAKORE,CEKR}.

The critical value of $\lambda$ shifts very slowly to higher values
for increasing $N$, although the increase is not
statistically significant.

In addition the shape of the specific heat (for example the width)
is basically
unchanged as we go to larger lattices. From Figs. 2 and 3 we do not infer
evidence of criticality.

In Fig. 4a we show the radius of gyration of the surface, $R_G$, as defined
in (\ref{E_RG}). Here obviously the volume scaling is non-trivial: larger
surfaces have larger radius. The histogram reconstruction already ceases
to work for quite low values of $\lambda$ for the larger lattice.
This effect could be related to the interesting finite size scaling behavior
of this quantity, which we illustrate in better detail in Fig. 4b.
Here we plot

\be
  \nu(N) \equiv \frac{\log \frac{R(N)}{R(\frac{N}{2})}} {\log(2)}\ .
\ee

This is an effective inverse Hausdorff dimension, which is a function of
$\lambda$. In the large $\lambda$ limit $\nu \to 1$ and $d_{extr}\to 2$, as
expected for flat surfaces. In the low $\lambda$ limit $d_{extr}$ becomes
very large. In the pseudo-critical region $\nu$ is a linear function of
$\lambda$. Curiously enough, the latter curve yields a Hausdorff
dimension of $4$, a value characteristic of branched polymers, near the
location of the specific heat peak.
This value is not particularly reliable though because of finite-size effects
and a value changing rapidly from $2$ to a large number must pass through
$4$ somewhere in the crossover region.  In
ref. \cite{AMBTWO} a value compatible with ours ($D_H(\lambda_c)>3.4$)
is quoted for the critical theory.
We stress however (and also here we are in complete agreement with
\cite{AMBTWO}) that the Hausdorff dimension in the pseudo-critical
region depends heavily and quite unusually on $N$.

In both the high and low $\lambda$ regions finite size effects are
quite small (compatible with zero to one standard deviation).
In the pseudo-critical region, on the contrary,
finite size effects are large. This effect
cannot be explained by the shift in $\lambda$ which one gets from the shift of
the peak of the specific heat, which is far too small. This behaviour is very
different from that we discussed for $S_E$ and it seems to
indicate the possibility of some sort
of critical behavior close to $\lambda=1.4$.

In Fig. 5 we plot the expectation values of the magnitude of the extrinsic
Gaussian curvature  $\mid {\cal K} \mid$. If the induced metric is equal
to the intrinsic metric, then $\mid {\cal K} \mid =
\frac{\mid {\cal R} \mid}{2}$.

This plot is not substantially
different from that of $S_E$. We note that
finite size effects are somewhat larger in this case than for
the edge action, but they follow the same pattern (exhibiting a
big increase in the flat
phase).

The plot of the fluctuations of the extrinsic Gaussian curvature,
$F[{\cal{K}}]$, which we
present in Fig. 6, shows
something very new. A very sharp crossover, with perhaps a
peak developing for large $N$, dominates the pseudo-critical behavior.
Fluctuations do not seem to depend on $\lambda$ in the crumpled phase, while
they drop dramatically, in a very small $\lambda$ interval, in the flat
region. Here again, finite size effects are sizeable in the pseudo-critical
region. The position of the crossover does not depend sensitively
on $N$, while the
detailed shape at $\lambda_c$ seems to change slightly with $N$.

It is difficult to give a precise interpretation of a plot like this, but,
as we said, the crossover is very clear here.

In Fig. 7 we give the intrinsic curvature ${\cal{R}}$ and in Fig. 8 its
fluctuations. Both plots are very similar to the related,
extrinsic curvature,
$\cal{K}$ plots.
$\mid {\cal{R}} \mid$ drops off rapidly, just as $\mid {\cal{K}} \mid$
does.
Through the peak region, though, $\mid {\cal{K}} \mid$ decreases
by about a factor
of $5$ while $\mid {\cal{R}} \mid$ diminishes to only about $.6$ of its value
on the left-hand side of the peak.  Since the action explicitly suppresses
mean curvature, and the mean and extrinsic
Gaussian curvature are closely related
(for instance, $H^2 > \frac{K}{2}$), we would expect that for large
$\lambda$  extrinsic fluctuations would be suppressed much more than
fluctuations of intrinsic geometry.

In Fig. 9 we plot the intrinsic extrinsic curvature correlation. The plot of
${\cal K}*{\cal R}$ indicates that intrinsic and extrinsic geometry are
strongly correlated for small $\lambda$, but as one passes through the peak
region they become decorrelated.  This is not particularly surprising, given
that the action directly suppresses only extrinsic fluctuations.

In Fig. 10a we plot the expectation value of the maximum coordination number,
which has non-trivial scaling behavior. In Fig. 10b we give its scaling
exponent, defined analogously to the exponent we have exhibited for the
gyration
radius.  In the
pseudo-critical region $q_{max}$ scales (for our $3$ lattice sizes) as a
power, with an exponent close to $0.1$; we do not know if this scaling
is meaningful.

\section{Correlation Times \protect\label{S_TIM} }

We will discuss here correlation times for different observables. As we
already pointed out correlation times become very large in the large $\lambda$
region. In agreement with ref. \cite{AMBTWO} (see their Fig. 1) we do not see
any increase of the correlation times close to the pseudo-critical point.

We will not present precise estimates of correlation times (exponential or
integrated) -- they are too large to get precise estimates. We will limit
ourselves to a discussion of a few figures, which give quite a clear idea of
what is happening. The comparison with Fig. 1 of ref. \cite{AMBTWO} cannot be
very direct, since our action is different, and because their dynamics
may be more effective than ours. Still, the comparison is quite
puzzling, since we estimate and exhibit correlation times which are
much (orders of magnitude)
larger than the ones of \cite{AMBTWO}.  Applying customary
methods to estimate $\tau_{int}$ can lead to an underestimate of short
correlation times
if more than one time scale is present (that does surely happen to our data
if we integrate our data on a window of reasonable size).

In Fig. 11a we plot $S_E$ for $N=144$, $\lambda=1.4$, and in Fig. 11b
the gyration
radius for these values (with a different time scale). Clearly, the
correlation time is at least of order
$40,000$ sweeps in the first case and $100,000$ sweeps in the
second one. In Figs. 12a and b, we plot the same quantities for $\lambda=1.5$.
Here correlation times are larger, of order $50,000$ steps for $S_E$ and
larger than $150,000$ steps for $R_G$. In Figs. 13 we draw the same plot on
the largest lattice we study ($N=576$) for $\lambda=1.4$. Here we can see
dramatic correlations, with times of
at least $100,000$ steps for $S_E$ and of at
least $1,000,000$ steps for $R_G$.

In Fig. 14 we plot, for the same time history and on the same scale, both
$S_E$ and $R_G$. This figure shows a clear anticorrelation: larger surfaces
are flatter and have smaller curvature (this is apparent in the
region close
to the 2000th step).

\section{Interpretation of Results \protect\label{S_RES} }

The results of this paper show that the
transition from crumpled to flat surfaces with increasing
$\lambda$ is quite complex. Guided by the present data
and the very interesting results of ref. \cite{AMBTWO}
we present in this section our current interpretation
of the situation.

On lattices with up to $576$ vertices we can clearly see a sharp
crossover, but the absence of a diverging specific heat, of diverging
correlation times and of strong finite size effects suggests that we are not
observing a usual second order phase transition. On the other hand,
quantities like the intrinsic curvature, or the mass measurements of
\cite{AMBTWO} show that something non-trivial is happening.

Let us review the crux of our observations again.
This model of crumpled surfaces appears to
exhibit sharp crossover behavior in the region around $\lambda = 1.4$.  The
sharp growth in the gyration radius and the suppression of curvature
fluctuations indicate that the normals acquire long-range correlations, up to
the size of the systems we examine.   Presumably the zero string tension
measurement of
\cite{AMBTWO} also shows that the disordered regime differs from the regime in
which the surfaces are ordered (up to scale of the lattices that are
simulated) by only  a small shift in $\lambda$. This evidence  might indicate
the presence of a phase transition at this point.  If so, it is very likely
to be of order higher than $2$ (or, rather implausibly, it
could be second order
with an extremely low negative specific heat exponent; our lattices are
much too small for us to confidently extrapolate the value of the specific
heat exponent as $N \rightarrow \infty$).

If the transition were higher order, the peak should exhibit a cusp, but we
would need far more accurate data to detect this.  The existence of this phase
transition would then suggest the existence of a new continuum string theory,
though many other issues would have to be resolved (e.g. unitarity) to
determine if such a theory is physically desirable.

There are other possible interpretations of our data.  We  need to
consider the influence of finite-size effects, since  the surfaces which we
simulate are quite small, even smaller than one might naively assume because
they are not intrinsically smooth.  For instance, random surfaces
characteristic of $D=0$ gravity have a Hausdorff dimension of roughly
$d_{intr} = 2.8$ \cite{KAWNIM,Migtalk}; it has been predicted that surfaces
embedded in $1$ dimension have Hausdorff dimension $2 + \sqrt{2}$
\cite{KAWNIM}.  Thus, for instance, if the surfaces in our simulations had an
intrinsic dimension of $3$, they would have a linear size of fewer than $9$
lattice spacings\footnote{Of course, our lattices are too small,  by one or
two orders of magnitude, to really exhibit a convincing fractal structure.}.

Perhaps the simplest alternative explanation  for the presence of this peak is
suggested by the arguments of Kroll and Gompper \cite{KROGOM}.  They argue
that the peak occurs when the persistence length of the system approaches the
size of the lattice ($\xi_p \sim N^{\frac{1}{d}}$)\footnote{For couplings
above this point, our simulations would simply be measuring finite size
effects.}. For larger $\lambda$, fluctuations on a larger scale become more
important, but when this scale is greater than the lattice size, these
fluctuations are suppressed. Thus one might surmise that the specific heat
will drop for large $\lambda$. (It clearly goes to zero for small $\lambda$
because of the presence of the prefactor $\lambda^2$; the lattice
implements a ultraviolet cutoff that freezes out very short-range
fluctuations.)

The one-loop renormalization group calculation (\ref{persist}) predicts that
the persistence length grows as $\xi_p \sim \exp(C\lambda)$; $C$ is inversely
proportional to the leading coefficient of the  beta function.  We would
expect that the peak position should shift  to the right with increasing $N$
in this scenario as

\begin{equation}
  \Delta =
  \frac{\delta\ln N}{d_{intr}C}.
\end{equation}

Quite a large value of $C$ is needed to explain the  rapid crossover; roughly
values of $C \sim 10, d_{intr} \sim 3$ are more or less consistent with the
magnitude of the peak shift and crossover width.  The RG calculations using
different forms of the action yield $C = \frac{4\pi}{3}$ (see equation
\ref{persist}), but
this may not apply to the action we simulate.

This reasoning also indicates that the peak should widen as the lattice size
increases; we do not observe this at all. It seems plausible though that
these arguments, based only on the leading term of the high lambda expansion,
are too naive.

      An alternative scenario, which builds on the ideas in the above
paragraph, is suggested by the tantalizing similarities between the results of
our fluid surface simulations and what has been observed for the $d=4$ $SU(2)$
Lattice Gauge Theory \cite{ENZOHI} and for the $d=2$ $O(3)$ model. Let us
discuss the case of the $O(3)$ model.

The $O(3)$ model, which is asymptotically free, exhibits a specific heat peak
near $\beta = 1.4$.  This peak was first measured via Monte Carlo
simulations by Colot \cite{Colot}. It can also be obtained by differentiating
the energy data measured by Shenker and Tobochnik \cite{ShTob,Brout}.  The
origin of this peak is understood \cite{Brout,Brout2}; it is due to the
fluctuations of the sigma particle, a low-mass bound state of the massless
$O(3)$ pions.  The sigma induces short-range order, and contributes to the
specific heat as a degree of freedom only at high temperatures (when the
correlation length in the system becomes smaller than its inverse mass).
The peak thus occurs at the beginning of the crossover regime, when the
correlation length is several lattice spacings.

According to the low temperature expansion, the correlation  length grows as
$\xi \sim \exp(2\pi\beta)/\beta$. Thus one would expect  a fairly rapid
crossover in the $O(3)$ model;  the correlation length should increase by
roughly a factor of $9$ when $\beta$ is shifted by about $.35$\footnote{In
fact,  the presence of the sigma significantly modifies this  low-temperature
expansion result \cite{Brout2}  in this intermediate regime, but does not
qualitatively destroy the rapidity of the crossover.  Indeed, despite heroic
efforts, it has been impossible to extend computationally beyond this regime
and
precisely verify the asymptotic low-temperature relation for the correlation
length \cite{Apost,Hasen}.}.  Such a crossover is indeed observed, though it
is not so apparent that it is as dramatic as the crossover behaviour
observed for fluid surfaces.\footnote{To quantitatively
compare the width of the crossover regimes for these two models it would
be necessary to measure a correlation length (perhaps
extracted from the normal-normal correlation function) in these random
surface simulations.}

Recent simulations of the $O(3)$ model \cite{Kostas} indicate that the
specific heat peak grows significantly when the lattice size $L$ is increased
from $5$ to $15$, and that virtually no growth in peak height is evident as
$L$ is increased further up to $100$.  Also, the peak position shifts to the
right as $L$ grows, and then appears to stabilize for large $L$.  This is more
or less what we observe in our simulations of fluid surfaces, on lattices of
small size.  We point out these similarities largely to emphasize that there
does exist an asymptotically free theory (with low mass excitations) which
exhibits crossover behavior qualitatively similar to that
observed in our simulations.

The analogy is perhaps deeper, though, since the fluid surface action (with
extrinsic curvature) in certain guises looks like a sigma model action.  So,
perhaps it would not be so surprising from this point of view to find a sigma
particle in these theories perhaps associated with ($\hat{n}^2 -1$), in which
$\hat{n}$ denotes the unit normal to our surfaces.

Another additional possibility is  that fluctuations of the intrinsic geometry
(the Liouville mode) are  responsible for short-range order and contribute to
the specific heat peak.

\section{Further Work \protect\label{S_WHA} }

There remains much to be done to clarify whether or not a crumpling transition
occurs for a finite value of the extrinsic curvature coupling $\lambda$.  It
would be interesting (and probably a fair amount of work) to apply Wilson
renormalization group techniques to the actual action (\ref{ourcontact}) which
we simulate, to determine the leading coefficient of the beta function.
Additionally, perhaps  a calculation of $1/D$ corrections to the large $D$
computations already performed could unearth evidence of a sigma-type
excitation in these theories (the effects of the sigma appear as $1/N$
corrections in the $O(N)$ model).

       We also are histogramming our data to examine the behavior of complex
zeroes (in complex $\lambda$ space) of the partition function of our
simulations \cite{winprog}.   It has been shown (in the case of $SU(2)$
lattice gauge theory) that such zeroes, when they are near but do not approach
the real axis in the infinite volume limit, occur in theories which exhibit
specific heat peaks with no associated phase transition \cite{ENZOHI}. Low
temperature expansions also indicate that the $O(3)$ model susceptibility
has a complex singularity near the real axis \cite{Butera}-- presumably this
corresponds to a zero of the partition function and is a manifestation of the
sigma.

Of course, simulations on large lattices, with better statistics, should also
help us evaluate whether a crumpling transition exists.  We are testing
algorithms, such  as simulated tempering \cite{simtemp}, in order to evade the
long  auto-correlation times that have characterized our simulations so far.

      Even if no such transition exists for finite $\lambda$, one
could still attempt to study a continuum theory in the strong coupling
limit, as is done for QCD, for instance.  To do so, we would like
to examine global quantities, such as masses extracted from normal-normal
correlation functions, rather than just the local quantities (energy, e.g.)
that we have measured.  Measuring these correlations requires a definition
of distance on these triangulated lattices; the most successful definition
of the metric is based on the propagation of massive particles (via inversion
of the Laplacian) on these lattices \cite{MigLap}.
\section{Conclusion}
   We have thus explored the phase diagram of fluid random surfaces
with extrinsic curvature, but unfortunately we have been unable to determine
if our model undergoes a phase (crumpling) transition at finite coupling.
We have observed dramatic crossover behavior for particular observables
in our Monte Carlo simulations, but on the other hand, the correlation times
and certain finite-size effects do not behave as one would expect in the
presence of a phase transition.
The behavior of other lattice models also indicates that it is possible
that we are observing the effects of finite-mass excitations on small
lattices, rather than a phase transition.  We hope that future work
will clarify this murky state of affairs, to determine if there indeed exists
a crumpling transition for fluid surfaces.

\section*{Acknowledgments}

This work has been done with NPAC (Northeast Parallel Architecture
Center) and CRPC (Center for Research in Parallel Computing)
computing facilities,
which we thank. The research of MB was supported by the Department
of Energy Outstanding Junior Investigator Grant DOE DE-FG02-85ER40231
and that of GH by research funds from Syracuse University.

We gratefully acknowledge discussions, help and
sympathy from Jan Ambj{\o}rn,
Kostas Anagnostopoulos, John Apostolakis, Clive Baillie, Mike Douglas,
David Edelsohn, Volyosha Kazakov,
Emil Martinec, Alexander Migdal, David Nelson,
Giorgio Parisi, Bengt Petersso Steve Shenker and Roy Williams.

During the first stage of this project we used the DIME code (written by
Clive Baillie,  Desmond Johnston and Roy Williams)
to get acquainted with simulations of random surfaces.
Paul Coddington, Leping Han, and Geoffrey Harris would like to
thank especially Clive Baillie and Roy Williams for their
assistance with this code.  Mark Bowick
and Geoffrey Harris would also like to acknowledge the
hospitality of the Department of Physics, Universit\'a di
di Roma {\em Tor Vergata},
and with Enzo Marinari, we would also like to thank the Department
of Theoretical Physics, Ecole Normale Sup\'erieure, where some of this
work was done.

Geoffrey Fox
has had an important role in this project, both through interesting
discussions and collaboration on domain decomposition of
dynamical triangulated meshes and also through his general support. We are very
grateful to him for that.

The entire NPAC organization
has been very helpful.
We are especially indebted
to Deborah Jones for organizing all that needed to be organized,
and to Peter Crockett, Mark Levinson, and Nancy McCracken and many others
for their wonderful computational support.

\vfill
\newpage

\section{Figure Captions}
  \begin{itemize}

    \item[Fig. 1] The edge curvature $S_E$ as a function of $\lambda$.
As in all other pictures, filled circles and a dotted line correspond to
$N=144$, crosses and
a dashed line indicate $N=288$, and empty squares and a solid line
represent $N=576$.

    \item[Fig. 2] The edge curvature specific heat, $C(\lambda)$.

    \item[Fig. 3] As in Fig. 2, but with the multi-histogram
    reconstruction in
the transient region.

    \item[Fig. 4a] The gyration radius $R_G$ defined in
(\ref{E_RG}), plotted as in Fig. 1.

    \item[Fig. 4b] The effective inverse Hausdorff dimension $\nu$ as a
function of $\lambda$, as defined in (\ref{E_RGD}). The filled dots and the
dashed curve are from a fit to the $N=288$ and $N=144$ data, while the empty
dots and solid curve represent the fit to $N=576$ and $N=288$.

    \item[Fig. 5] The extrinsic Gaussian curvature
$\mid{\cal{K}}\mid$ defined in (\ref{E_CK}), plotted as in Fig. 1.

    \item[Fig. 6] The fluctuations of $\mid {\cal{K}} \mid$.

    \item[Fig. 7] The  intrinsic curvature
$\mid{\cal{R}}\mid$ defined in (\ref{E_CR}), plotted as in Fig.1.

    \item[Fig. 8] The fluctuations of $\mid {\cal{R}} \mid$.

    \item[Fig. 9] The intrinsic extrinsic curvature
correlation, as defined in (\ref{E_KR}), plotted as in Fig.1.

    \item[Fig. 10a] The average maximum coordination
number of the surface vertices,  ${\rm{max}}_i \ q_i$, plotted as in
Fig.1.

    \item[Fig. 10b] The scaling exponent of
${\rm{max}}_i\ q_i$, plotted as in Fig.1.

    \item[Fig. 11a] $S_E$ as a function of Monte Carlo time ($80,000$
steps) for $N=144$, $\lambda=1.4$.

    \item[Fig. 11b] $R$ as a function of Monte Carlo time ($300,000$
steps) for $N=144$, $\lambda=1.4$.

    \item[Fig. 12a] $S_E$ as a function of Monte Carlo time ($80,000$
steps) for $N=144$, $\lambda=1.5$.

    \item[Fig. 12b] $R$ as a function of Monte Carlo time ($300,000$
steps) for $N=144$, $\lambda=1.5$.

    \item[Fig. 13a] $S_E$ as a function of Monte Carlo time ($300,000$
steps) for $N=576$, $\lambda=1.4$.

    \item[Fig. 13b] $R$ as a function of Monte Carlo time ($3,000,000$
steps) for $N=576$, $\lambda=1.4$.

    \item[Fig. 14] $S_E$ and $R$ from the same Monte Carlo run, $N=576$,
$\lambda=1.325$, $20,000$ steps.

\end{itemize}

\vfill

\end{document}